\documentclass[fp,a4paper]{jpsj3}

\usepackage{mathrsfs}
\usepackage{txfonts}
\def\tc{$T_{\textrm c}$}   
\def\kca{K$_2$Cr$_3$As$_3$} \def\korb{$K_{\textrm{orb}}$}
\def\kchi{$K$-$\chi$ plot} \def\hct{$H_{\textrm{c2}}$}

\def\dvec{$\textbf{d}$}
\usepackage[dvipdfmx]{graphicx}

\title{Single Crystal Growth of and Hyperfine Couplings in the Spin-Triplet Superconductor K$_2$Cr$_3$As$_3$}

\author{Seigo Ogawa$^1$, Tomoki Miyoshi$^1$, Kazuaki Matano$^1$, Shinji Kawasaki$^1$, Yoshihiko Inada$^2$, and Guo-qing Zheng$^1$}
\inst{$^1$Department of Physics, Okayama University, Okayama 700-8530, Japan \\
$^2$Faculty of Education, Okayama University, Okayama 700-8530, Japan} %\\

\abst{
We report single crystal growth of the spin-triplet superconductor  K$_2$Cr$_3$As$_3$ with superconducting transition temperature $T_{\textrm c}$ = 6.2 K,
and the measurements of magnetic susceptibility $\chi$ as well as the Knight shift above $T_{\textrm c}$. 
We determined the hyperfine coupling constants directly from the relation between the Knight shift ($K$) and susceptibility ($K$-$\chi$ plot) and obtained the orbital contribution  $K_{\textrm{orb}}$. 
Our results of $K_{\textrm{orb}}$ is in fairly good agreement with the previous estimate by a novel method. Using the obtained parameters, we  calculated the spin susceptibility $\chi_{\textrm{s}}$, which is found to be  isotropic above {\tc}, %indicating that  spin-orbit interaction is small in this system.
in spite of a quasi-one dimensional crystal structure.}
\begin{document}
\maketitle
\thispagestyle{plain}
\pagestyle{plain}
\section{Introduction}
More than a century after the discovery of superconductivity, study on it is still thriving.
In the long history of research, countless superconductors have been discovered and their properties revealed.
Among them, spin-triplet superconductivity has attracted particular attention,  %been found in only a few cases.
%\cite{UPt3_PhysRevLett.52.679,UPT3_Tou_PhysRevLett.80.3129,Togano_PhysRevLett.93.247004,Nishiyama_PhysRevLett.98.047002,Hor_PhysRevLett.104.057001,MatanoKrienerSegawaEtAl2016}.
because of its rarity and remarkable physical properties and application perspectives \cite{MatanoKrienerSegawaEtAl2016,zheng}. %ed interest, the elucidation of the mechanism of triplet superconductivity and the search for new materials have been vigorously pursued.
%Furthermore, the fact that it can be a topological superconductor with Majorana fermions on the surface and inside the vortex threads has further increased its attention.
%
%In recent years, {\kca} has attracted attention as a possible member of the spi-triplet superconductor\cite{Bao2015,Yang2021}.
{\kca} is a recently discovered superconductor that contains the 3d transition metal Cr\cite{Bao2015}.
It was expected that magnetism  would play an important role in the occurrence of superconductivity, as in other 3d transition-metal compound superconductors such as high-$T_c$ copper oxides \cite{Bednorz}, cobalt oxides \cite{Takada} and iron pnictides \cite{Hosono}.
Subsequently, it was found that K can be replaced by other alkali elements which resulted in the discovery of a series of superconductors  Rb$_2$Cr$_3$As$_3$\cite{Rb2Cr3As3_PhysRevB.91.020506}, Cs$_2$Cr$_3$As$_3$\cite{Cs2Cr3As3_Tang2015}, and Na$_2$Cr$_3$As$_3$\cite{Na2Cr3As3_Mu2018IonexchangeSA}. % have been discovered in which .
Unlike other transition-metal compound superconductors, which have apparent two-dimensional(2D) crystal structures\cite{Bednorz,Takada,Hosono}, 
this system has a seemly quasi-one-dimensional crystal structure where two Cr$_3$ chains run along the $c$-axis. \cite{Bao2015}
%The one-dimensional structure of $A_2$Cr$_3$As$_3$($A$=Na,K,Cs,Rb) consists of Cr$_3$As$_3$ linear chains strung in the c-axis.
The  inner Cr$_3$ chains and outer As$_3$ chains are %, with Cr$_3$As$_3$
 separated from each other by alkali metals.
Evidence of unconventional superconductivity have been reported.
No coherence peak was observed below {\tc} in the temperature dependence of the spin-lattice relaxation rate $1/T_1$ \cite{K2Cr3As3_NMR_Zhi2015,NMR_FerromagneticSF_Yang2015} which showed a power-law behavior ($1/T_1\propto T^5$) at low temperatures, suggesting point nodes in the gap function \cite{NMR_FerromagneticSF_Yang2015,A2Cr3As3_Luo2019}.
The penetration depth measurements \cite{lambda_PhysRevB.91.220502,uSR_PhysRevB.92.134505}, 
and the electronic specific heat %was proportional to $T^2$ 
 \cite{Shao_2018} also show a power-law behavior at low temperatures.
The upper critical field {\hct} exceeds the Pauli limit\cite{Hc2_PhysRevB.95.014502},
and the {\tc} decreases significantly with a small amount of impurities\cite{impurity_Liu2016}.
In the normal state above $T_c$,
nuclear quadrupole resonance (NQR) measurements found  ferromagnetic spin correlations \cite{NMR_FerromagneticSF_Yang2015,A2Cr3As3_Luo2019}, 
 in contrast to antiferromagnetic spin correlations in copper oxides \cite{Millis}, iron-based superconductors\cite{Oka_FeAs_PhysRevLett.108.047001}, and cobalt oxides\cite{Fujimoto,Matano_NaxCoO2_EPL}.
%It has been pointed out that magnetic fluctuations may be responsible for superconductivity, as in copper oxides and iron-based superconductors \cite{theory_Wu2015MagnetismIQ}.
%Furthermore, in $A_2$Cr$_3$As$_3$, 
The relationship between the ferromagnetic fluctuation and  spin-triplet superconductivity has been studied both theoretically and experimentally \cite{Wu_theory_PhysRevB.92.104511,A2Cr3As3_Luo2019}.  %in $A_2$Cr$_3$As$_3$($A$=Na,K,Cs,Rb) has attracted much attention.
%Frustration originating from the triangular lattice of Cr is proposed to be the cause of magnetic fluctuations\cite{Theory_frustration_PhysRevB.103.214406},
%Ferromagnetic spin fluctuation is condidered favorble for spin-triplet superconductivity

Recently, the Knight shift $K$ was measured in a single crystal {\kca} by $^{75}$As nuclear magnetic resonance (NMR) \cite{Yang2021}. The Knight shift is the most direct probe for detecting the paring symmetry.
It was found that $K$ does not decrease below {\tc} with the magnetic field applied perpendicular to the $c$-axis, but decreases  with the magnetic field parallel to the $c$-axis.
%It is also expected to be a strong candidate for topological superconductors\cite{Theory_Topological_Xu2020}.
%Among the signs of unconventional superconductivity, 
%However, it would require a more precise discussion since some argue for spin-singlet superconductivity
%\cite{singlet_Hc2_Balakirev2015,singlet_theory_PhysRevB.99.094511}.
%There remains some issue with this Knight shift measurement, and that is the determination of {\korb}. 
The Knight shift $K$ consists of two parts as $K=K_{\textrm{s}}+K_{\textrm{orb}}$, where $K_{\textrm{s}}$ is due to spin susceptibility
and $K_{\textrm{orb}}$ is due to orbital susceptibility which is $T$-independent.
Determination of {\korb} is essential if one wants to know how much spin susceptibility changes below {\tc}.
%That is, the reduction in the Knight shift is meaningful only if the orbital part of the night shift is accurately determined.
In the previous study, {\korb} was determined using a novel method, namely, from the relation between the Knight shift and the spin-lattice relaxation time $T_1$. It was concluded that in both field directions, $K_{\textrm{s}}$ is substantially large which decreases toward zero only for $H\parallel$ $c$-axis.
Thus, the result indicates spin-triplet superconductivity with the vector order parameter ({\dvec}-vector) oriented along the $c$-axis\cite{Yang2021}.
 
In the previous study, however, the hyperfine couplings between nuclear and electron spins was not determined.
This knowledge is important since it allows one to extract the spin susceptibility that can be used to estimate various physical quantities and the detailed properties of the superconducting phase(s).
%This is usually in good agreement with usual metals, but care must be taken in the case of this system. This is because spin fluctuations, if present, will affect $T_1$.
%Furthermore, the presence of ferromagnetic spin correlations has been noted in this system. 
%While antiferromagnetic spin correlations have a negligible effect on the Knight shift\cite{Matano_NaxCoO2_EPL,Matano_BaFeAs}, ferromagnetic correlations do affect the Knight shift. 
%Therefore, it is not appropriate to estimate {\korb} using the Knight shift and $T_1$ in this system.
In this paper, we report on the synthesis of single crystals of {\kca}, sample characterization and the determination of hyperfine coupling constants.
%We have successfully obtained high-quality single crystals.
Using the Knight shift obtained in this study for $H\parallel c$ and the data for $H\perp c$ reported in Ref \cite{Yang2021}, and the dc magnetic susceptibility data obtained in this work, we performed the {\kchi}.
We determined the orbital part of the Knight shift \textit{K}$_{\rm{orb}}$ more directly, and extract the spin susceptibility.
%Hyperfine interaction constants in each direction of the axes were determined.

\section{Experimental Methods}

\subsection{Sample Synthesis}

The single crystals of K$_2$Cr$_3$As$_3$ were grown by using a high temperature solution growth method\cite{Canfield}. 
Potassium pieces, chromium powder, and arsenic granules were packed in an alumina crucible at a molar ratio of K:Cr:As = 6:1:7 following Ref \cite{Bao2015}. 
The alumina crucible and the materials were sealed in a tantalum crucible under an argon atmosphere. 
Then, the tantalum crucible was sealed with a quartz tube. They were heated up to 1000\(^\circ\) C and kept at 1000\(^\circ\) C for two days to allow a good mixing of  the starting materials. After cooling down to 700\(^\circ\) C at a speed of less than 3 K/h, the single crystals and remaining liquid were separated in a centrifuge.
We extended the time reported in  Ref.\cite{Canfield} to allow raw materials to meld well.
We obtained about 30 single crystals per batch.

\subsection{Measurements}
We coated samples with Apiezon N grease to prevent exposure to air and packed a few needle-shaped single crystals in a bundle. We observed the superconducting transition by measuring 
the temperature dependence of dc magnetic susceptibility at \textit{H} = 10 Oe applied perpendicular to the $c$-axis, using a Quantum Design Magnetic Property Measurement System (MPMS) for both zero-field-cooling (ZFC) and field-cooling (FC).
In addition, we measured dc susceptibility of the samples in the normal state (7 K $\leq$ \textit{T} $\leq$ 300 K) at \textit{H} = 3 T for \textit{H} $\parallel$ \textit{c} axis and \textit{H} $\perp$ \textit{c} axis, respectively.
%In any measurements, we used a few samples coated with Apiezon N grease to prevent exposure to air.
$^{75}$As NMR measurements were carried out by using a phase-coherent spectrometer. The NMR spectra were obtained by scanning the frequency point by point and integrating the spin echo at a fixed magnetic field. In calculating the Knight shift, the nuclear gyromagnetic ratio $\gamma$ = 7.2919 MHz/T was used. 
The procedure to obtain the Knight shift is the same as Ref.\cite{Yang2021}.

\section{Results and Discussions}

\subsection{Superconducting Transition and Magnetic Susceptibility in the Normal State}
\begin{figure}
	\centering
	\includegraphics[width=8cm]{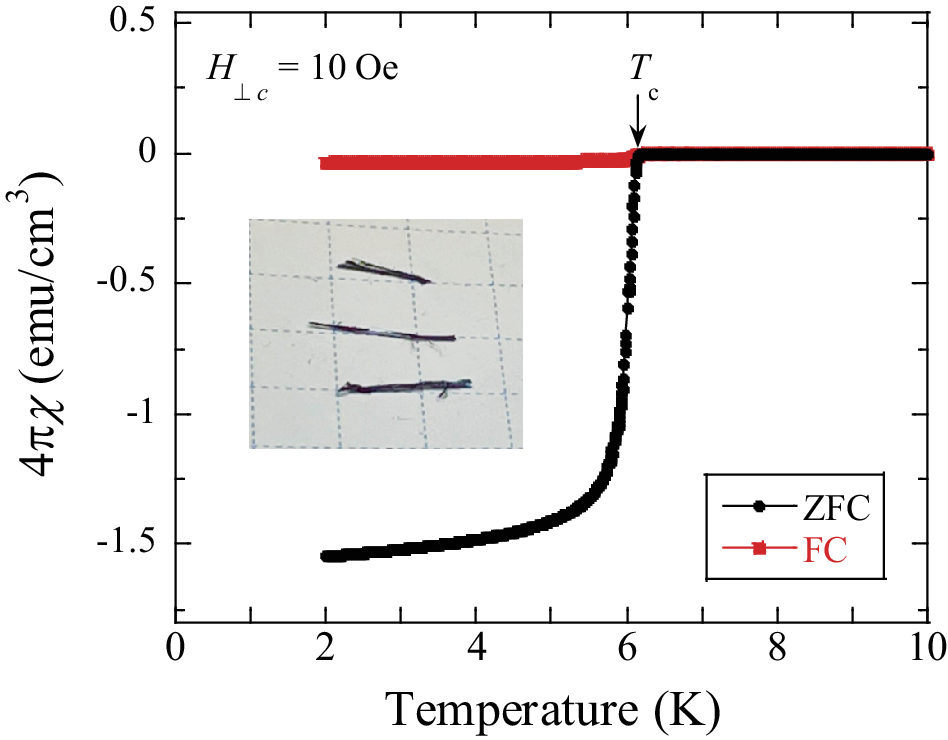}
	\caption{(color online) Temperature dependence of dc magnetic susceptibility around {\tc}. The inset shows the harvested single crystals of K$_2$Cr$_3$As$_3$ on the 5mm grid paper.}
	\label{dcchi}
\end{figure}

Figure \ref{dcchi} shows the temperature dependence of dc magnetic susceptibility for ZFC and FC, respectively. 
The samples exhibit diamagnetism below \textit{T}$\rm{_c}$ $\simeq$ 6.2 K in ZFC, and the superconducting volume fraction  is sufficiently large to claim bulk superconductivity.  
%exceed 100\%. 
\begin{figure}
	\centering
	\includegraphics[width=8cm]{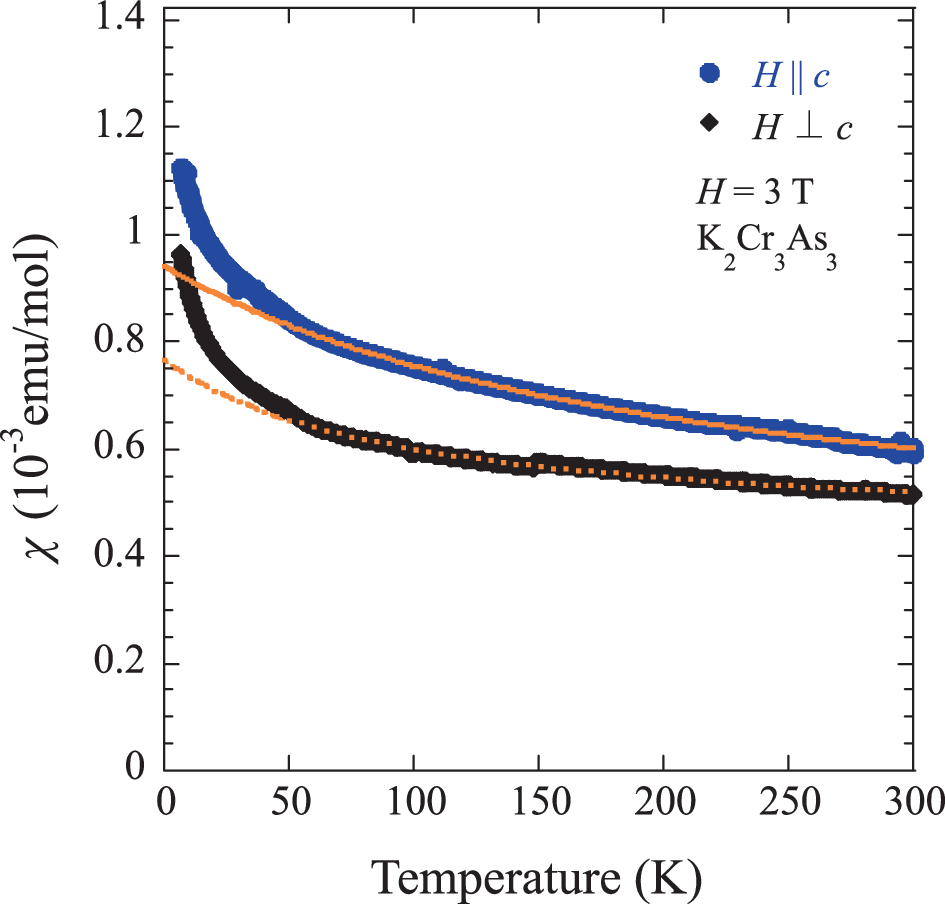}
	\caption{(color online) Temperature dependence of dc magnetic susceptibility of K$_2$Cr$_3$As$_3$ measured at  \textit{H} = 3 T for \textit{H} $\parallel$ \textit{c} axis and  \textit{H} $\perp$ \textit{c}, respectively. The solid and dashed curves show fittings of the data above 50 K with a Curie-Weiss relation $\chi = \chi_0+ \frac{C}{T+\theta}$. 
		The obtained parameters are $\chi_0$ = 6.31×10$^{-4}$ (emu/mol) and $C$ = 215 K for $H\parallel c$, and $\chi_0$ = 7.14×10$^{-4}$ (emu/mol)  and $C$ = 97 K for $H\perp c$, respectively.}
		%Insets show temperature dependence of the Knight shift for both directions reported previously \cite{Yang2021}.}
	\label{dcchi2}
\end{figure}
Figure \ref{dcchi2} shows the temperature dependence of dc susceptibility.
The results are in good agreement with previous reports\cite{Bao2015,Canfield}.

\subsection{NMR Spectra and the Knight Shift}
Figure 3 shows the NMR spectra for the central transition ($m$ = +1/2 to -1/2 transition) at a field about 13 T (to be precise, $H$ = 12.951 T) applied parallel to the $c$-axis. The full width at half maximum (FWHM) for the As2 site is 50 kHz, which is smaller than 100 kHz
 previously obtained at 12 T \cite{Yang2021}, suggesting good quality of our crystal.
Figure \ref{kshift} shows the temperature dependence of the Knight shift, which is  in good agreement with the previous data \cite{Yang2021}.
\begin{figure}
	\centering
	\includegraphics[width=8cm]{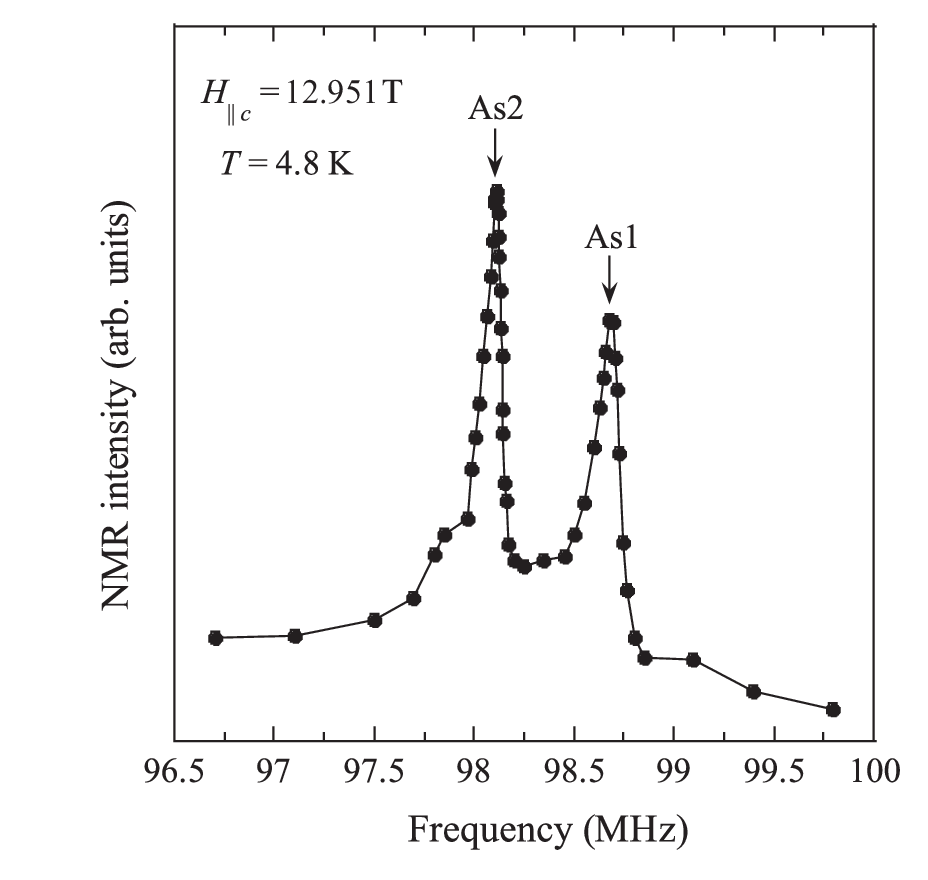}
	\caption{$^{75}$As-NMR spectra for the central transition taken at a field of $\sim$ 13 T applied parallel to $c$-axis.}
	\label{spec}
\end{figure}

\begin{figure}
	\centering
	\includegraphics[width=8cm]{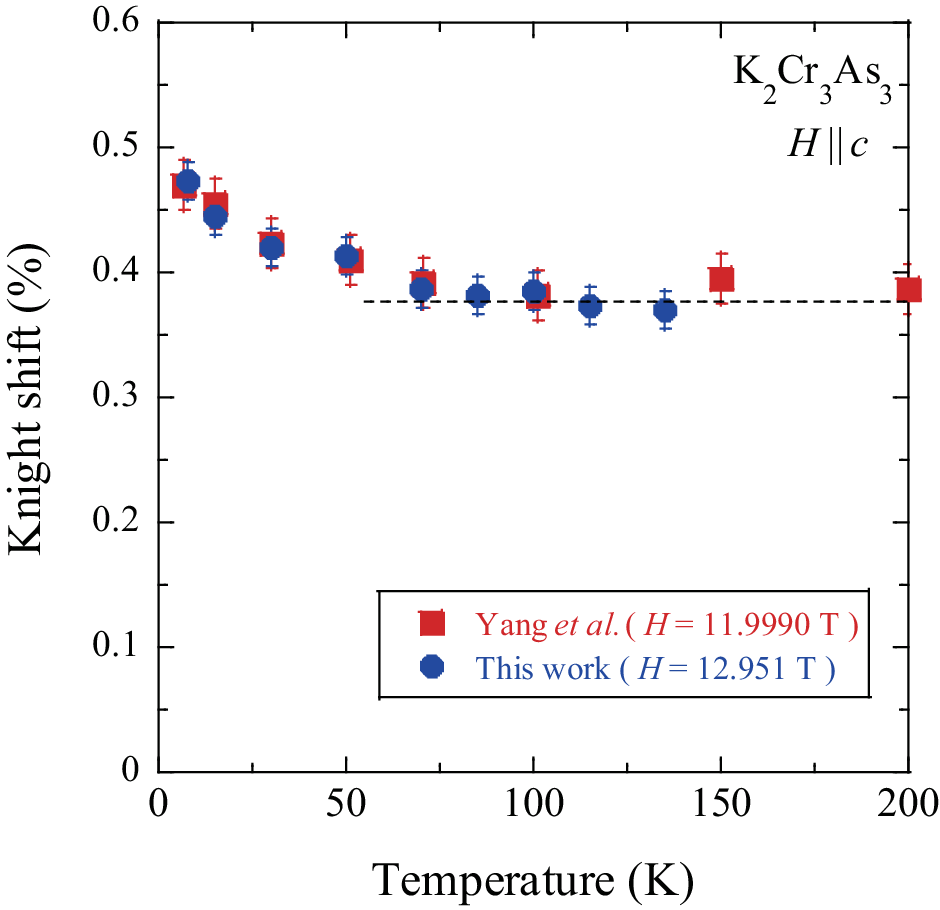}
	\caption{(color online) Temperature dependence of the Knight shift at the As2 site above {\tc} compared with the previously reported data. The horizontal broken line is the guide to the eyes.}
	\label{kshift}
\end{figure}

\subsection{Determination of the Hyperfine Coupling Constants and the Orbital Part of the Knight Shift}

The Knight shift is expressed as follows,
\begin{eqnarray}
	%\chi\!\!&=&\!\!\chi\rm{_s} + \chi\rm{_{orb}} +  \chi\rm{_{core}},\\
	K\!\!&=&\!\!K\rm{_s} + \textit{K}\rm{_{orb}}\\
	K\rm{_s}\!\!&=&\!\!\textit{A}\rm{_{hf} }\,\chi\rm{_s}\\
	K\rm{_{orb}}\!\!&=&\!\!\textit{A}\rm{_{orb}}\,\chi\rm{_{orb}}\, = \,2\,\genfrac{<}{>}{1pt}{}{1}{\textit{r}^3}\,\chi\rm{_{orb}}
\end{eqnarray}
where \textit{K}$_{\rm{s}}$ is temperature dependent and is related to the spin susceptibility $\chi_{\textrm{s}}$ via the hyperfine coupling constant \textit{A}$_{\rm{hf}}$.
{\korb} is the contribution from orbital susceptibility which is $T$-independent.
%\textit{K}$_{\rm{dia}}$ is the contribution from vortex lattice formation in the superconducting state. 
%According to the calculation\cite{Yang2021}, the  \textit{K}$_{\rm{dia}}$ is negligible in K$_2$Cr$_3$As$_3$. 
%The hyperfine coupling constants  for \textit{H} $\parallel$ \textit{c} axis and \textit{H} $\perp$ \textit{c} axis  were estimated from the slope of  \textit{K} - $\chi$ (emu/mol) plot to be 
The Knight shift appears to become a constant above $T$ = 50 K, while
the dc susceptibility shows a weak temperature dependence even at high temperatures which is attributable to a contamination of  magnetic impurities possibly coming from  remaining flux. %This is an open question
 %This indicates that above and Knight shift, the dc susceptibility is temperature dependent even in the range where .
%As shown in Equation 2, since the Knight shift is proportional to the dc susceptibility, the dc susceptibility should also be constant in the region where the Knight shift is constant.
%This may be due to the presence of a small amount of Curie-Weiss-like impurities in the sample.
In performing the {\kchi}, we tentatively subtract the high-temperature contribution to the dc susceptibility as follows.
The dc susceptibility above 50 K was fitted by a  Curie-Weiss relation,  % to remove the effects of impurities. 
%The fitting was done using the Curie-Weiss equation expressed as follows:
$\chi = \chi_0+ \frac{C}{T+\theta}$,
as shown by the curves in Fig. 2.
By subtracting the term $\frac{C}{T+\theta}$, we remove the impurity contribution to $\chi$. % Using obtained parameters, the effect of Curie-Weiss impurities was subtracted from the dc susceptibility.

\begin{figure*}[t]
	\centering
	\includegraphics[width=16cm]{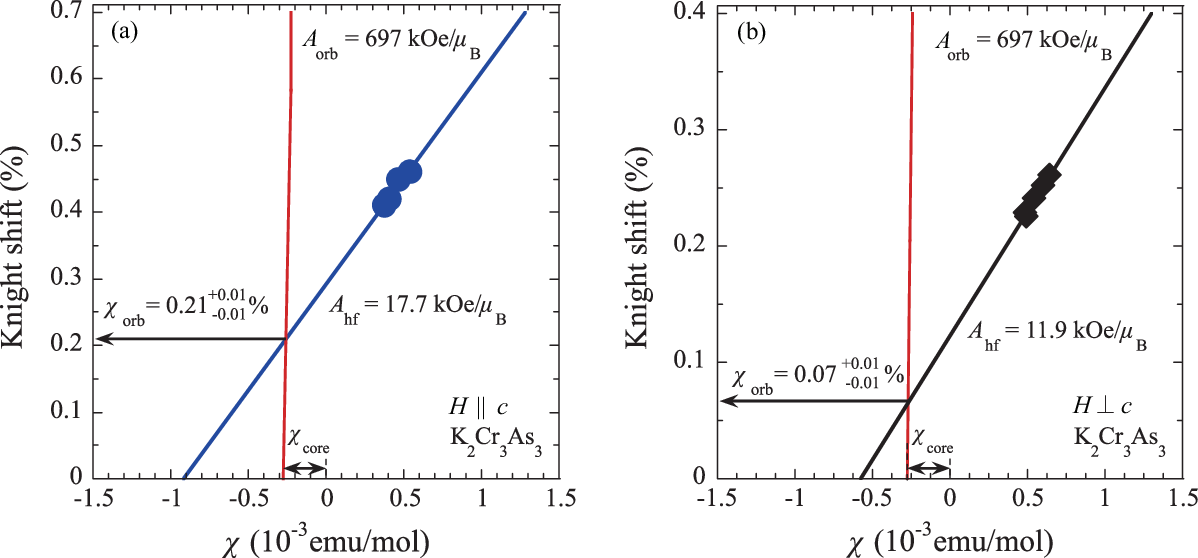}
	\caption{(color online) {\kchi} for (a) \textit{H}  $\parallel$ \textit{c}-axis and (b)  \textit{H} $\perp$ \textit{c}-axis, respectively. The $K$ data for \textit{H} $\perp$ \textit{c} were taken from Ref.\cite{Yang2021}.}
	\label{f3}
\end{figure*}

\begin{table*}[hbtp]
	\caption{Hyperfine coupling constants and the orbital contributions to the susceptibility and the Knight shift.} %Comparison with the results reported in Ref \cite{Yang2021}.}
	\label{t1}
	\centering
	\scalebox{2}{}
	\begin{tabular}{lcccccc}
		\hline\hline
		& $A_{\rm{hf}}^ c$(kOe/$\mu$$_B$) & $A_{\rm{hf}}^ {\perp c}$(kOe/$\mu$$_B$) & $K_{\rm{orb}}^ c$(\%) &  $K_{\rm{orb}}^{\perp\,c}$(\%)& $\chi_{\rm{orb}}^ c$(emu/mol)  &  $\chi_{\rm{orb}}^{\perp\,c}$(emu/mol)  \\
		\hline%hline
		Yang \textit{et al}.  & -  & - &  $0.27,(+0.01/ -0.02)$  & $0.09,(+0.01 / -0.02)$&-&- \\
		This work  & 17.7  & 11.9 &  $0.21,(+0.01 / -0.01)$  & $0.07,(+0.01 / -0.01)$&1.7$\times$10$^{-5}$&5.6$\times$10$^{-6}$   \\
		\hline\hline
	\end{tabular}
\end{table*}

\begin{figure}
	\centering
	\includegraphics[width=8cm]{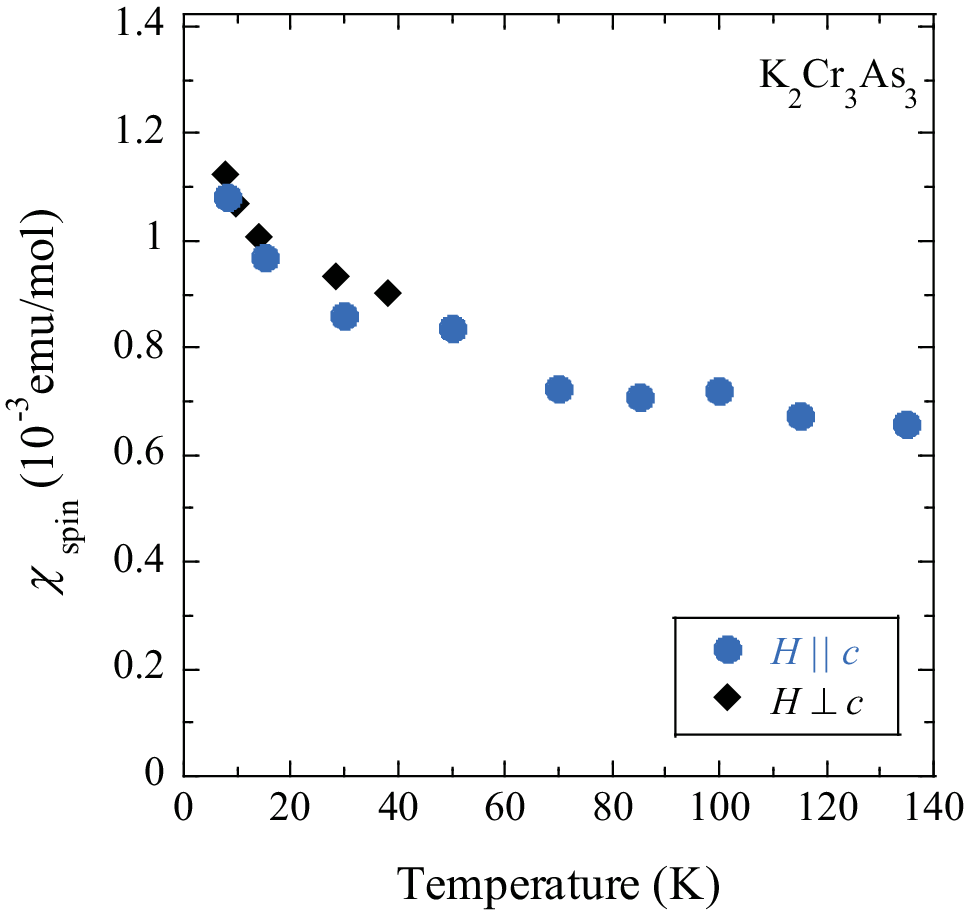}
	\caption{(color online) Temperature dependence of the spin susceptibility {$\chi_{\textit{s}}$} above {\tc}.}
	\label{f3}
\end{figure}
We then perform the {\kchi} plot to estimate the hyperfine coupling constant and  \textit{K}$_{\rm{orb}}$ in each direction, using the normal-state Knight-shift data for $T\leq$ 50 K. As can be seen  in Fig. 5, a good linear relation between $K$ and $\chi$ is obtained, from which the hyperfine coupling constant was obtained as $A_{\rm{hf}}^ c$ = 17.7 kOe/$\mu$$_B$, and \textit{A}$_{\rm{hf}}^{\perp c}$ = 11.9 kOe/$\mu$$_B$.
%To perform K-chi for As nuclei, the units are changed from emu/mol to emu/mol-As. mol-As is calculated from the ratio of the atomic weight of AsA to that of As.
The contribution of   \textit{K}$_{\rm{orb}}$ is estimated by the following standard method.
In addition to $\chi_{\rm s}$ and $\chi$$_{\rm{orb}}$, 
there is also a contribution from  diamagnetic susceptibility due to  closed shells of K, Cr, As, so that $\chi$ is written as $\chi$=$\chi_{\rm s}$+ $\chi$$_{\rm{orb}}$ + $\chi$$_{\rm{core}}$. The $\chi$$_{\rm{core}}$ is
 calculated to be $\chi$$_{\rm{core}}$ = - %1.34$\times$10$^{-4}$ (emu/mol-As)\cite{handbook}.
 2.74$\times$10$^{-4}$ (emu/mol)\cite{handbook}.
 In Fig. 5, a straight line representing the relation
 \textit{K}$_{\rm{orb}}$ = \textit{A}$_{\rm{orb}}$$\,$$\chi$$_{\rm{orb}}$ was drawn from the horizontal-axis position which is shifted from  the origin by the amount of  $\chi$$_{\rm{core}}$.
Here, $\langle$1/\textit{r}$^3$$\rangle$$_{\rm{As}}$ = 5.57 a.u. is adopted, which is 80\% of the theoretical value for As metals\cite{atomic}.  
%The error of \textit{K}$\rm{_{orb}}$ was estimated from the slope and intercept in drawing the line by the least squares method.
We thus obtained  $K_{\rm{orb}}^ c = 0.21\%\ \pm 0.01\% $ and %/ -0.03\%)$,
 $K_{\rm{orb}}^{\perp c} = 0.07\% \ \pm 0.01\% $.  %,(+0.01\% / -0.02\%)$.
  The  error is the   standard deviation.
  As can be seen in Table 1, the obtained {\korb} is in fairly good agreement with the previous estimate \cite{Yang2021}.
   %for \textit{K}$\rm{_{orb}}$ was estimated as follows. The lower bound was estimated by omitting the data points below 55 K in the fittings of Fig. 2. On the other hand, including more data points down to 30 K does not affect the estimation of \textit{K}$\rm{_{orb}}$.  Therefore, the standard deviation of the fittings in Fig. 3 was taken as the upper bound. For completeness,  we  also performed the  calculation without subtracting the impurity contribution.
%In that case, {\korb} was calculated to be $K_{\rm{orb}}^ c =0.18\%$ and $K_{\rm{orb}}^{\perp c} = 0.05\%$, respectively.

Figure 6 shows the temperature dependence of the spin susceptibility  $\chi_{\rm s}$ calculated by using the Knight shift data obtained in this work for $H \parallel c$ and the data of Ref.\cite{Yang2021} for $H \perp c$, and the $A_{\rm{hf}}$ and $K_{\rm{orb}}$ obtained above.
%The $T^*$ is the temperature at which the Knight shift begins to decrease.
%{\kca} has multiple phases in the superconducting state and $T^*$ corresponds to the phase boundary\cite{Yang2021}.
The upturn of $\chi_{\rm s}$ at low temperatures is due to ferromagnetic spin correlations\cite{NMR_FerromagneticSF_Yang2015,A2Cr3As3_Luo2019}.
The $\chi_{\rm s}$ is isotropic above {\tc}, which is usually the case when the spin-orbit coupling is small.
A large spin-orbit coupling could lead to an anisotropy in spin susceptibility \cite{Eremin,LiZ}. Also, the result is consistent with the total  DOS being dominated by  the 3D Fermi surface \cite{LDA_Jiang2015}, in spite of a quasi-one dimensional crystal structure.
Our results indicate that the previous conclusion is justified. Namely, 
the spin susceptibility $\chi_{\rm s}$ decreased below {\tc} when the magnetic field is applied parallel to $c$ axis, while it did not change when the field is perpendicular to $c$ axis.
This clearly indicates that {\kca} is a spin-triplet superconductor.
For spin-singlet superconductivity, the $\chi_{\rm s}$ should decrease by the same amount under a magnetic field perpendicular to the $c$-axis as the case when the magnetic field is applied parallel to the $c$-axis.

%Our analysis confirms that {\kca} is still a spin-triplet superconductor.
%The dc susceptibility used for{\kchi} was calculated by subtracting the effect of impurities, 
%The error of {\korb} are the larger of the error derived from the calculation of the slope of $K\rm{_s}=\textit{A}\rm{_{hf} }\chi\rm{_s}$ or the difference from the case where the effect of impurities is not subtracted.
%

\subsection{Summary}
We have successfully grown single crystals of {\kca} and observed a large superconducting volume fraction that ensures bulk superconductivity. We measured the magnetic susceptibility for both $H \parallel c$ and $H \perp c$ and  the Knight shift for $H \parallel c$.  We derived the hyperfine coupling constants and  {\korb} directly from the {\kchi}. The obtained  {\korb} is in fairly good agreement with the value estimated previously. %, which indicates that the spin susceptibility is substantially large in both field direction.
Using the obtained parameters, we  calculated $\chi_{\textrm{s}}$ for each direction, and found that $\chi_{\textrm{s}}$ is isotropic above {\tc}, in spite of a quasi-one dimensional crystal structure. This suggests that %as expected for a system with a small
the spin-orbit coupling is small. These results indicate that  the previous conclusion of {\kca} being a spin-triplet superconductor is justified, as the Knight shift results indeed imply a nematic response of $\chi_{\textrm{s}}$ below {\tc}.
%The orbital part of the Knight shift {\korb} and hyperfine coupling constant {\hyp} were determined by {\kchi}.
 %, but exhibits a different behavior below {\tc}.
%The behavior of $\chi_{\textrm{s}}$ provides strong evidence to reinforce that {\kca} is a spin-triplet superconductor.

\begin{acknowledgment}
We thank D. Aoki for the help in making the Tantalum crucible, and  T. Kambe for  help in the susceptibility measurements.
This work was supported in part by the JSPS  Grants No. 19H00657, No. 20K03862 and No. 22H04482 (Grant-in-Aid for Scientific Research on
Innovative Areas “Quantum Liquid Crystals”).
\end{acknowledgment}

\bibliographystyle{JPSJ}  

\bibliography{k-chi_plot_paper}

\end{document}